\documentclass[fleqn,usenatbib]{rasti}
\usepackage{newtxtext,newtxmath}
\usepackage[T1]{fontenc}
\DeclareRobustCommand{\VAN}[3]{#2}
\let\VANthebibliography\thebibliography
\def\thebibliography{\DeclareRobustCommand{\VAN}[3]{##3}\VANthebibliography}

\usepackage{graphicx}	
\usepackage{amsmath}	
\usepackage{xcolor}
\definecolor{myred}{rgb}{0.67571825, 0, 0.17578125}
\definecolor{myblue}{rgb}{0, 0.203125, 0.3828125}
\usepackage{hyperref}
\hypersetup{
    colorlinks=true,
    linkcolor=myblue,
    urlcolor=myblue,
    citecolor=myblue,
    filecolor=myblue
}

\newcommand\Nall{1,458,666}
\newcommand\Nunusable{236}
\newcommand\Nusable{1,458,430}
\newcommand\Ncand{48,814}

\newcommand\Nfinalnom{402}
\newcommand\NMBAnom{390}
\newcommand\NNEAnom{13}

\newcommand\Nfinalext{2144}
\newcommand\NMBAext{1971}
\newcommand\NNEAext{86}
\newcommand\NOtherext{87}

\newcommand\Nthree{357}
\newcommand\Nseven{209}
\newcommand\Nthirty{49}

\newcommand\HV{$H_{\rm V}$~}

\title[Arial small bodies]{Spectroscopic Survey of Small Solar System Bodies with Ariel}
\author[J. Beniyama et al.]{
Jin Beniyama$^{1,2}$\thanks{E-mail: jbeniyama@oca.eu (OCA)},
Eri Tatsumi$^{3,4,2}$,
Sunao Hasegawa$^{5}$,
Keigo Enya$^{5}$,
Matt Griffin$^{6}$,
\newauthor
Andreas Papageorgiou$^{6}$,
Lorenzo V. Mugnai$^{6}$,
Csaba Kiss$^{7,8,9}$,
Moe Matsuoka$^{10}$,
Takafumi Ootsubo$^{11}$,
\newauthor
and Subhajit Sarkar$^{6}$\\
$^{1}$Université Côte d'Azur, Observatoire de la Côte d'Azur, CNRS, Laboratoire Lagrange, Bd de l'Observatoire, CS 34229, 06304 Nice Cedex 4, France \\
$^{2}$ Department of Earth and Planetary Science, The University of Tokyo, 7-3-1 Hongo, Bunkyo, Tokyo 113-0033, Japan\\
$^{3}$Instituto de Astrofísica de Canarias, C/ V\'{i}a L\'{a}ctea, s/n, San Cristobal de La Laguna, Tenerife, Spain\\
$^{4}$ Departamento de Astrof\'{i}sica, Universidad de La Laguna, Avenida Astrofísico Francisco Sánchez, s/n, San Cristobal de La Laguna, Tenerife, Spain\\
$^{5}$ Institute of Space and Astronautical Science, Japan Aerospace Exploration Agency, 3-1-1 Yoshinodai,
Chuo-ku, Sagamihara, Kanagawa 252-5210, Japan\\
$^{6}$Department of Physics and Astronomy, School of Physical, Chemical and Environmental Sciences, Cardiff University, The Parade, Cardiff CF24 3AA, UK\\
$^{7}$Konkoly Observatory, HUN-REN Research Centre for Astronomy and Earth Sciences, Konkoly Thege Miklós út 15-17, H-1121
Budapest, Hungary\\
$^{8}$CSFK, MTA Centre of Excellence, Budapest, Konkoly Thege Miklós út 15-17, H-1121, Hungary\\
$^{9}$ELTE Eötvös Loránd University, Institute of Physics and Astronomy, Pázmány Péter sétany 1/A, Budapest, Hungary\\
$^{10}$National Institute of Advanced Industrial Science and Technology\\
$^{11}$University of Occupational and Environmental Health,  1-1 Iseigaoka, Yahatanishi, Kitakyusyu, Fukuoka 807-8555, Japan
}
\date{Accepted XXX. Received YYY; in original form August 07, 2026}

\pubyear{\the\year{}}
\begin{document}
\label{firstpage}
\pagerange{\pageref{firstpage}--\pageref{lastpage}}
\maketitle

\begin{abstract}
The small Solar System bodies (SSSBs) preserve records of the formation and evolution of the Solar System. In particular, the near-infrared wavelength region around 3 $\mu$m provides crucial insights into the origin and transport of hydrated materials and organics, although it is difficult to observe from the ground because of strong telluric absorption.
We assess the observability of SSSBs as complementary science to the ESA Ariel space mission, operating in the visible to mid-infrared wavelength range (0.5--7.8~$\mu$m). Using the NASA JPL Horizons ephemeris system, we evaluate the visibility of known SSSBs from 2032 to 2035. 
Our analysis shows that Ariel can spectroscopically characterize up to 402~unique SSSBs under a conservative limiting visual magnitude of $V=13.6$. If observations can be extended to somewhat fainter targets ($V=16$), the number of observable SSSBs could increase to 2144.
Assuming operational scenarios in which one SSSB is observed every 3, 7, or 30 days over a four-year mission, the total numbers of observable targets would be 357, 209, and 49, respectively. We further model the spectral energy distributions of candidate targets to evaluate their observational performance. We demonstrate that Ariel can resolve critical absorption features, such as those at 2.7~$\mu$m and 3.4~$\mu$m, with a higher spectral resolution than the previous infrared survey with AKARI. Observations of SSSBs with Ariel can provide insights into the distribution and evolution of life-forming materials in the Solar System and
constitute a scientifically compelling component of Ariel’s complementary science program.
\end{abstract}


\section{Introduction}
The small Solar System bodies (SSSBs) record the history of the Solar System in their physical properties such as orbits, rotations, reflectance spectra, and thermal properties.
Thus, studying SSSBs gives clues to understanding our planet's formation as well as entire planetary systems.
Among the characteristics of SSSBs, the near-infrared (NIR) reflectance spectra provide crucial information in the context of the origin of water and life \citep[e.g.,][]{Campins2010}. 
For example, the 2.7-$\mu$m feature is linked to hydrated minerals, the 3.1-$\mu$m feature is related to water ice, and the 3.4-$\mu$m feature is related to organics \citep[e.g., ][]{Takir2012, Rivkin2015, Usui2019}.
Furthermore, testing interpretations and hypotheses based on ground-based visible-to-near-infrared spectral surveys up to 2.5 $\mu$m \citep[e.g.,][]{Hasegawa2026} requires a wider wavelength coverage.
It is not impossible to characterize features around 3~$\mu$m from the ground when the asteroid is bright enough.
The NIR spectra including 3 $\mu$m (hereafter referred to as the 3-$\mu$m spectra) of SSSBs have been extensively investigated using the NASA Infrared Telescope Facility ($\approx$0.8--4.2~$\mu$m) in numerous studies \citep[see, e.g.,][]{Rivkin1995, Rivkin2010, Campins2010, Takir2012, Takir2020}. However, it is challenging to characterize around these wavelengths from the ground due to the telluric absorptions, especially at 2.7--2.9 $\mu$m, and higher-quality data can be obtained from space.


The first 3-$\mu$m spectra of asteroids from space were obtained using the Infrared Space Observatory \citep[ISO;][]{Kessler1996}.
\citet{Rivkin1997} presented the 3-$\mu$m spectra of (1) Ceres and (2) Pallas across the 2.5--3.5~$\mu$m wavelength range, revealing distinct absorption features.
One of the most successful surveys of the 3-$\mu$m spectra of asteroids from space was performed with the Japanese space satellite AKARI \citep{Murakami2007}.
The AKARI Spectroscopic Observations (Aqua-spec) consists of 147 pointing observations of SSSBs, and obtained NIR spectra of 66 unique SSSBs from 2.5 to 5~$\mu$m \citep{Usui2019}. 
The targets were primarily selected from main belt asteroids (MBAs), Cybeles, Hildas with the exception of one near-Earth Asteroid (NEA) (4015) Wilson-Harrington.
Recently, 3-$\mu$m spectra of SSSBs have been obtained with 
the NIRSpec instrument on the James Webb Space Telescope ($\approx$0.6--5.3~$\mu$m) \citep[JWST, see, e.g.,][]{GomezBarrientos2024, Jarmak2024, Wong2024, Thomas2025, Rivkin2025, Arredondo2025,LePivert-Jolivet2025}.
Moreover, NASA's Spectro-Photometer for the History of the Universe, Epoch of Reionization, and Ices Explorer \citep[SPHEREx,][]{Crill2020} has been serendipitously obtaining asteroid spectra from 0.75 to 5.0 $\mu$m \citep{Ivezic2022}.
Also, the Twinkle satellite \citep[0.5-4.5~$\mu$m,][]{Stotesbury2022} is planned to launch in 2029, and is expected to perform observations of Solar System objects \citep{Edwards2019a, Edwards2019b}.

The number of SSSBs with high-quality space-based NIR data including the 3-$\mu$m band has remained limited since AKARI.
Although JWST is expected to observe asteroids in the coming years, the number of samples will remain limited. 
In addition to the highly competitive telescope allocation process, its high sensitivity renders it unsuitable for observing bright targets.
SPHEREx is expected to provide spectra of SSSBs and may significantly advance our understanding of their compositions across various populations \citep[e.g.,][]{Lisse2025, Lisse2026, Lisse2026b}, although their calibration remains challenging due to the need for corrections regarding rotation and phase angle, as data at different wavelengths are acquired at different times \citep{Ivezic2022}.
To address this issue, simultaneous lightcurve observations of SSSBs observed by SPHEREx are being conducted as part of the Solar system Objects Light curve Observatory (SOLO) project \citep{Lim2026}.

Ariel (Atmospheric Remote-sensing Infrared Exoplanet Large-survey) is ESA's M4 mission scheduled for launch in 2031 \citep{Tinetti2018}. 
Its primary goal is to characterize the atmospheres of approximately 1,000 transiting exoplanets through transit and eclipse spectroscopy in the 0.5--7.8 $\mu$m range. 
The main spectroscopic instrument, AIRS (Ariel InfraRed Spectrometer), covers 1.95--7.8 $\mu$m in two channels (Ch0: 1.95--3.9 $\mu$m at R$\sim$100; Ch1: 3.9--7.8 $\mu$m at R$\sim$30--60) \citep{Eccleston2024}, providing access to key spectral features of silicates and organics relevant to primitive asteroids. 
Ariel also carries the Fine Guidance Sensor (FGS), the Visible Photometer (VISPhot), and the Near-IR Spectrometer (NIRSpec), which provide visible and near-infrared photometric and spectroscopic capabilities.
Although Ariel is not primarily designed for Solar System science, its infrared capability and observing cadence may offer opportunities for SSSB observations as part of its complementary science program. Using this opportunity, we can increase the NIR spectral samples of SSSBs and statistically discuss the composition of asteroids and their evolution through the early Solar System.


In this paper, we assess the observability of SSSBs with Ariel, with a particular focus on NIR spectroscopy using AIRS.
The paper is organized as follows.
Section~\ref{sec:method} describes our approach for assessing the observability of SSSBs from Ariel.
The results are presented in Section~ \ref{sec:res}.
The potential targets and expected scientific outcomes are discussed in Section~\ref{sec:disc}.

\section{Methods} \label{sec:method}
\subsection{Visibility}
We aim to assess the observability of SSSBs from Ariel at the Sun--Earth Lagrange point L$_2$.
We define an object as observable when its $V$-band apparent magnitude, $V$, is brighter than 13.6~mag for the nominal case or 16.0~mag for the extended case,
and its solar elongation lies between 70~deg and 120~deg.
The nominal limit corresponds to standard FGS tracking capabilities, on which current sensitivity estimates are based.
The extended limit accommodates fainter targets currently under evaluation for slow fine guidance mode, making its associated sensitivity performance preliminary
(see Section~\ref{subsec:SED} for details).
This elongation range is similar to, but slightly different from, that adopted by JWST (85~deg -- 135~deg). 
Because the stability limit on sky motion is one of the most uncertain parameters, we discuss it separately in Section 4. 

We first reduce the number of targets by filtering objects based on their orbital elements, in order to minimize the number of detailed ephemeris queries. We then use NASA JPL Horizons for accurate ephemeris calculations and evaluate their time-dependent apparent brightness.
To determine which small bodies are relevant for observations from Ariel at L$_2$, we estimate the lower limits of their $V$ band magnitudes under optimal conditions. 
We assume that Ariel is located at the L$_2$, Earth is at its aphelion, and each object is observed at perihelion, adopting opposition geometry and ignoring phase effects. 
These conditions yield the possible brightest apparent magnitude, $V_\mathrm{min}$.
This is a conservative assumption, and in principle, no targets would be missed unless their orbits become unstable within a few years.

We use the orbital elements catalog of SSSBs, \texttt{MPCORB} file\footnote{\url{https://www.minorplanetcenter.net/iau/MPCORB/MPCORB.DAT}, obtained on August 1, 2025.} available at the Minor Planet Center (MPC) website.
The \texttt{MPCORB} contains \Nall~SSSBs
\footnote{\
Here, SSSBs is used as a collective term that includes dwarf planets, although the official IAU (2006) definition of Small Solar System Bodies excludes them. Of the five currently recognized dwarf planets, 
Ceres and (134340)~Pluto are considered observable in this work.}.
We exclude \Nunusable~SSSBs without a value for \HV or with a value of 99.99,
which represents missing or unreported magnitude data, where \HV is the absolute magnitude in the $V$ band.
Thus, we use \Nusable~SSSBs.
We calculate the perihelion and aphelion distances of each object as $q = a(1-e)$ and $Q = a(1+e)$,
where $a$ and $e$ are the semi-major axis and orbital eccentricity, respectively.
We classify the SSSBs into three subpopulations:
near-Earth objects with $q < 1.3$ au, main-belt objects with $2.0 < a < 3.5$ au, and a residual population comprising all remaining objects.

The apparent magnitude in the $V$ band can be estimated as
\begin{equation}
V = H_V + 5 \log_{10} \left(r_{\rm obj} \, \Delta\right) - 2.5\log_{10}(\phi(\alpha)) \label{eq:Vmin}
,
\end{equation}
where $r_{\rm obj}$ is the heliocentric distance of the object,
$\alpha$ is the solar phase angle, and $\phi(\alpha)$ is the phase function normalized such that $\phi(0^\circ) = 1$.
The distance between Ariel and the object, $\Delta$, is given by
\begin{equation}
\Delta = \|\mathbf{r}_{\mathrm{Ariel}} - \mathbf{r}_{\mathrm{obj}}\|,
\end{equation}
where $\mathbf{r}_{\mathrm{Ariel}}$ and $\mathbf{r}_{\mathrm{obj}}$ are the heliocentric position vectors of Ariel and the object, respectively.
The heliocentric distance of Ariel varies slightly around the Sun--Earth L$_2$ point. 
Its maximum and minimum distances can be expressed as
\begin{align}
r_\mathrm{Ariel,max} &= Q_E + d_{L_{\rm 2,Sun\text{-}Earth}},\\
r_\mathrm{Ariel,min} &= q_E + d_{L_{\rm 2,Sun\text{-}Earth}},
\end{align}
where 
$Q_E = 1.0167$~au is the Earth's aphelion distance,
$q_E = 0.9833$~au is the Earth's perihelion distance,
and
$d_{L_{\rm 2,Sun\text{-}Earth}} \sim 0.01$ au is the Sun--Earth L$_2$ distance.

We classify objects into three categories based on the relation between their orbital range and Ariel's accessible heliocentric distance.
If the object remains entirely outside Ariel's maximum heliocentric distance ($q > r_\mathrm{Ariel,max}$), the minimum value of $r_{\rm obj}\,\Delta$ occurs when the object is at the perihelion, and Ariel is at $r_\mathrm{Ariel,max}$.
Assuming the ideal phase angle configuration ($\alpha \approx 0^\circ$, i.e., $\phi(\alpha) = 1$ at opposition) to evaluate the brightest theoretical limit, the minimum apparent magnitude is given by
\begin{equation}
V_\mathrm{min}
=
H_V
+
5 \log_{10}
\left(
q \,
\left|
r_\mathrm{Ariel,max} - q
\right|
\right).
\end{equation}
If the object remains entirely inside Ariel's orbit ($Q < r_\mathrm{Ariel, min}$), the minimum apparent magnitude occurs when the object is at aphelion and Ariel is at $r_\mathrm{Ariel,min}$.
Although an exact zero phase angle is geometrically unachievable in this configuration, assuming an ideal phase angle ($\alpha \approx 0^\circ$, i.e., $\phi(\alpha) = 1$) provides a conservative best-case upper bound on the brightness. The minimum apparent magnitude is thus given by
\begin{equation}
V_\mathrm{min}
=
H_V
+
5 \log_{10}
\left(
Q \,
\left|
r_\mathrm{Ariel,min} - Q
\right|
\right).
\end{equation}
If the object's orbit crosses Ariel's accessible heliocentric distance ($q < r_\mathrm{Ariel,max}$ and $Q > r_\mathrm{Ariel,min}$), the minimum value of $r_{\rm obj}\,\Delta$ depends on the detailed orbital geometry and cannot be determined analytically. In these cases, the minimum apparent magnitude $V_\mathrm{min}$ must be calculated using precise ephemerides.
This classification allows us to compute $V_\mathrm{min}$ analytically for most objects while restricting computationally expensive ephemeris queries only to intermediate cases, which are all NEAs (Apollos and Atens).

Objects for which this estimated $V_\mathrm{min}$ is fainter than the observational limit (i.e., $V = 16$) are excluded from further consideration, while the remaining objects are subsequently used to retrieve full ephemerides from JPL Horizons.
In total, \Ncand~objects, mostly NEAs, are considered to be potential targets.
We obtained the ephemerides with a time step of 1~day using the Python package \texttt{astroquery} \citep{Ginsburg2019}.
We set the \texttt{location} as \texttt{@gaia} to simulate observations from an L$_2$ orbit.
We query 
the right ascension (RA), 
declination (Dec), 
their associated
3$\sigma$ uncertainties ($\sigma_{\rm RA}$, $\sigma_{\rm Dec}$), 
ecliptic longitude ($\lambda$), ecliptic latitude ($\beta$),
heliocentric distance ($r_{\rm obj}$), observer-centric distance ($\Delta$), 
solar phase angle ($\alpha$),
$V$, 
solar elongation, 
and apparent velocity of SSSBs.

Additionally, we evaluated the positional uncertainty, defined as
$\sqrt{\sigma_{\rm{RA}}^2 + \sigma_{\rm{Dec}}^2}$,
and found that 33 objects have values larger than 5~arcsec.
We do not include these asteroids in the analysis.

\subsection{Spectral Energy Distribution} \label{subsec:SED}
To estimate the signal-to-noise ratio (SNR) across the wavelength range of Ariel, we model the Spectral Energy Distribution (SED) of an object.
The SED of SSSBs is expressed as the sum of the reflectance part, $F_{\rm ref}$, and the thermal emission part, $F_{\rm thermal}$.
The necessary parameters are extracted from various sources if available:
\HV from \texttt{MPCORB},
a geometric albedo in the $V$ band, $p_V$, and a spectral type
are retrieved from the \texttt{SsODNet} service
\citep{Berthier2023} through \texttt{rocks} Python interface.
Specifically, we adopted the best estimates for $p_V$ and the spectral type provided by \texttt{SsODNet.ssoCard}, the details of which are described in \citet{Berthier2023}.
The diameter, $D$, is calculated with $p_V$ and \HV using the following formula \citep{Fowler1992, Pravec2007}:
\begin{align}
D=\frac{1329}{\sqrt{p_\mathrm{V}}}\times10^{-H_{V}/5}. \label{eq:D}
\end{align}
Note that for Ceres and (433)~Eros, we adopted values of 
$D = 939.4$~km and $p_V = 0.09$ 
\citep[from NASA JPL Small-Body Database\footnote{\url{https://ssd.jpl.nasa.gov/tools/sbdb_lookup.html}}; see also][]{Li2006}, and $D = 16.84$~km and $p_V = 0.25$ \citep{Yeomans2000, Veverka2000}, respectively, instead of relying on the retrieved or calculated parameters.
The heliocentric distance and observer-centric distance at each epoch are queried from JPL Horizons as explained in the previous subsection.

The reflectance part of the SED is expressed as a scaled Solar spectrum as follows:
\begin{align}
F_{\rm ref}(\lambda) = S(\lambda) \cdot \left( \frac{F_V}{S_V} \right)  \cdot \frac{p(\lambda)}{p_V},
\end{align}
where 
$S(\lambda)$ is the Solar spectrum, 
$F_V$ is the asteroid's flux density at 0.55 $\mu$m calculated from the predicted $V$,
$S_V$ is the flux density of the Sun at 0.55 $\mu$m,
and $p(\lambda)$ is a wavelength-dependent geometric albedo.
We note that the phase function is included in the calculation of the predicted $V$.
For the purpose of this study, we assume that $S(\lambda) \sim B(\lambda, 5778\,K)$.
This assumption is justified because, in the subsequent analysis, the observed spectrum is normalized by the solar spectrum to remove the solar color and isolate the intrinsic reflectance properties of the target. 
In the actual analysis, the spectrum is therefore divided by a model spectrum in order to obtain the relative reflectance \citep{Usui2019}. 
Carbonaceous C-complex asteroids are known to have a similar SED to the Sun (i.e., G2V type stars), while silicaceous S-complex asteroids have redder SED than C-complex with absorption features around 1 and 2 $\mu$m.
To model these features, we use the spectral template from the latest Mahlke taxonomy \citep{Mahlke2022} and Bus-DeMeo taxonomy \citep{DeMeo2009} for $p_\lambda$ if the spectral type is available in the \texttt{SsODNet}. 
Otherwise, we assume C-type reflectance with the geometric albedo of 0.15.
For wavelengths longer than 2.5~$\mu$m, the reflectance is assumed to be constant, equal to the value at 2.5 $\mu$m in the template.
Furthermore, we assume the 3 $\mu$m band of Ceres and Pallas for C-complex asteroids \citep{Rivkin2025} in Section \ref{sec:disc}.

To model the thermal emission of the SSSB, we apply the simple Standard Thermal Model \citep[STM,][]{Lebofsky1986}.
This model was previously successfully utilized to derive asteroid sizes and albedos from the AKARI thermal infrared measurements \citep{Usui2011}.
The thermal flux is given by
\begin{align}
F_{\rm thermal}(\lambda) = \frac{\epsilon R^2}{\Delta^2} \int_{-\pi/2}^{\pi/2} \int_{-\pi/2 + \alpha}^{\pi/2} B(\lambda, T(\theta, \phi)) \cos^2 \theta \cos(\phi - \alpha) d\phi d\theta.
\end{align}
Here, 
$\epsilon$ is the emissivity, 
$R$ is the object radius, 
and $B(\lambda,T)$ is the Planck function.
We assume a constant, wavelength-independent value of $\epsilon = 0.9$.
The surface temperature distribution is expressed as
\begin{align}
T(\theta, \phi) = 
\begin{cases} 
T_{ss} \cos^{1/4} \theta \cos^{1/4} \phi & \text{(day)} \\
0 & \text{(night)} 
\end{cases}
\end{align}
, where $T_{ss}$ is the subsolar temperature given by
\begin{align}
T_{ss} = \left[ \frac{S_\odot (1 - A_{\rm Bond})}{\eta \epsilon \sigma r_h^2} \right]^{1/4}.
\end{align}
Here, $S_\odot$ is the solar constant at 1~au, 
$A_{\rm Bond}$ is the bolometric Bond albedo, 
$\eta$ is the beaming parameter, 
empirically introduced into the standard thermal model to account for surface roughness and the resulting sunward-directed thermal emission while compensating for inherent simplifying assumptions,
and $\sigma$ is the Stefan--Boltzmann constant.
We assume a constant value of $\eta = 0.756$.
The Bond albedo is given by
\begin{align}
A_{\rm Bond} = 1 - q(G)p_V, 
\end{align}
where $q(G)$ is the phase integral. 
We assume the phase integral $q(G=0.15)=0.4$ \citep{Bowell1989}.
Although the $G$ parameter is taxonomy-dependent, no strictly defined unique values exist for each class; thus, we adopt $G = 0.15$ as a standard baseline, subject to minor uncertainties in the phase angle correction.

To calculate the SNR of the observed spectrum, 
we use the sensitivity expressed in terms of the noise equivalent flux density (NEFD),
which is defined as the source flux density that yields an SNR of unity for an integration time of 0.5 s. 
The NEFD as a function of wavelength has been estimated using 
ArielRad \citep{Mugnai2020} using the current best estimates of the instrument characteristics.
In the NEFD calculation, detector noise, photon noise, pointing jitter noise, and system-gain noise are taken into account.
Using the NEFD, the $\rm{SNR}(\lambda)$ of the object with $F(\lambda)$ is calculated as follows: 
\begin{align}
{\rm SNR}(\lambda) = F(\lambda )\frac{\sqrt{2t_{\rm exp}}}{\rm NEFD},
\end{align}
where $t_{\rm exp}$ is an exposure time.

The sensitivity estimates adopted in this study apply only to targets brighter than approximately $V\sim13.6$, hereafter referred to as nominal candidates, 
which are sufficiently bright to be tracked using the nominal FGS mode. 
Targets fainter than this limit, hereafter referred to as extended candidates, may become observable using the proposed slow fine guidance mode, but this capability has not yet been confirmed. Furthermore, if such observations become feasible, increased pointing jitter may reduce the achievable sensitivity relative to the nominal FGS mode. 
Therefore, the results for extended candidates presented in this paper are intended for reference only and should not be interpreted as representative of the currently confirmed mission performance.

Since complementary science observations are typically executed within a 1--2 h observing block including telescope slewing and settling,
the individual exposure time should be limited to 10--15 minutes.
A nominal exposure time of 600~s is assumed throughout the paper.

By dividing the model SED by the model Sun's blackbody spectrum, 
we obtain the reflectance spectrum, $R(\lambda)$.
We assume the uncertainty is only from the asteroid spectrum. 
The useful wavelength range for reflectance should be carefully verified, 
as the assumption of a constant emissivity ($\epsilon = 0.9$) in the thermal correction could introduce ambiguities into the spectral features \citep[see][for details]{Usui2019}

\begin{figure}
	\includegraphics[width=\linewidth]{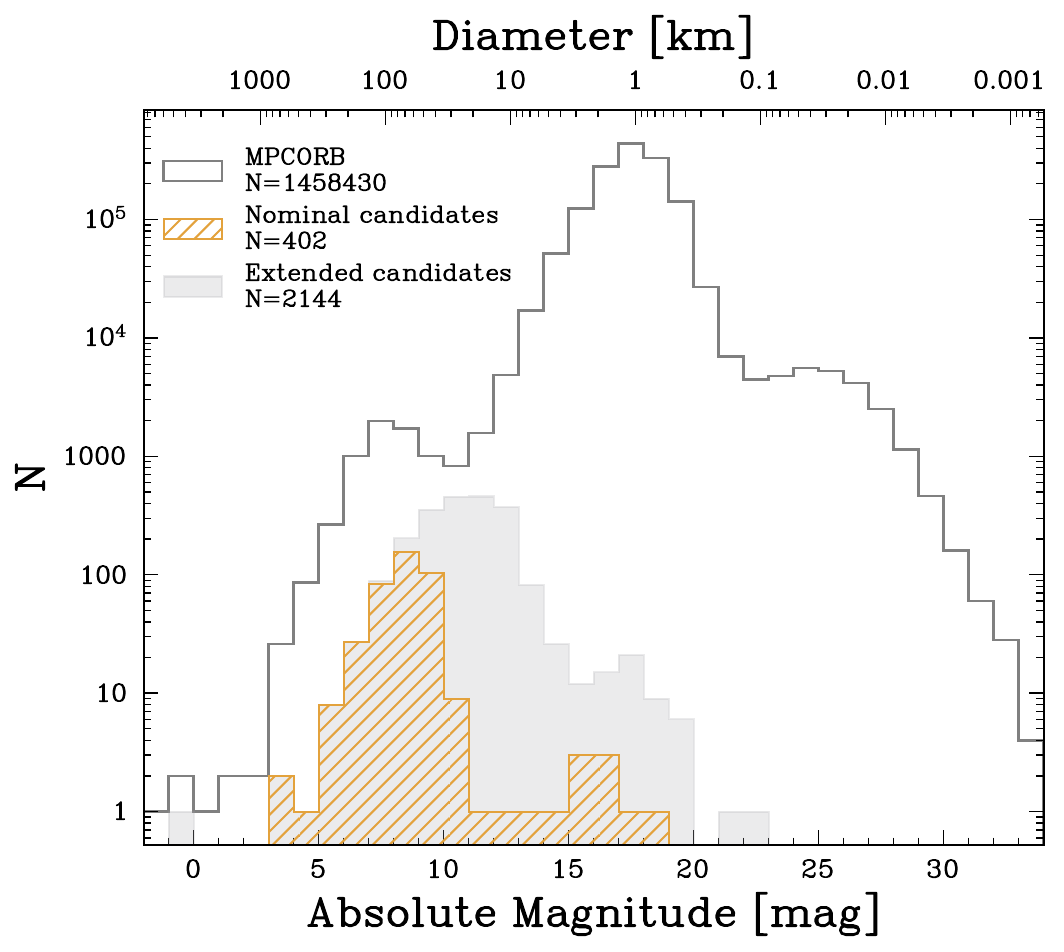}
    \caption{
    Histograms of 
    the SSSBs in the MPCORB catalog (solid line),
    the nominal candidates selected after a detailed ephemeris check with a criterion of $V < 13.6$ (striped pattern),
    and 
    the extended candidates selected after a detailed ephemeris check with a criterion of $V < 16.0$ (shaded region).
    The upper axis indicates the equivalent diameter assuming a geometric albedo of $p_V=0.15$.
    {Alt text: Histograms comparing the distributions of all cataloged and observable SSSBs.}
    }
    \label{fig:H_hist_1}
\end{figure}

\section{Results} \label{sec:res}
\subsection{Candidates}
\begin{figure}
	\includegraphics[width=\linewidth]{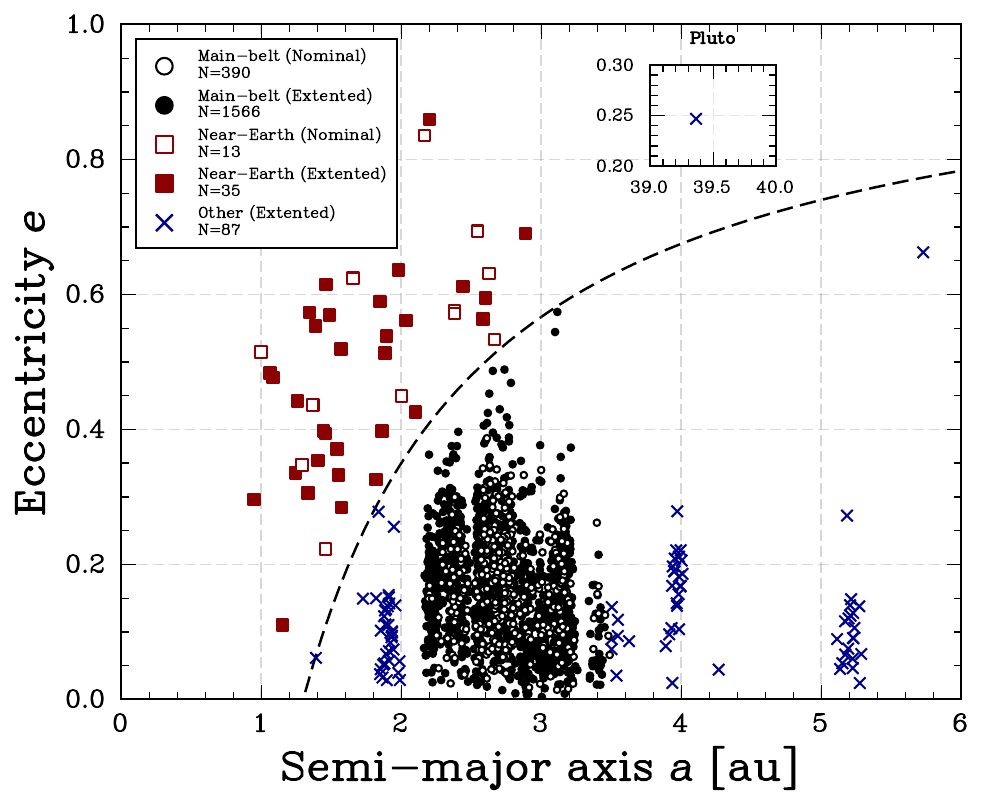}
    \caption{
    Distribution of the potential candidates in the semi-major axis--eccentricity plane.
    Main-belt objects, near-Earth objects, and others are represented by circles, squares, and crosses, respectively.
    Open symbols indicate the nominal candidates selected after a detailed ephemeris check with a criterion of $V < 13.6$, whereas filled symbols show the extended candidates selected with a criterion of $V < 16.0$.
    The dashed line indicates the boundary corresponding to a perihelion distance of $q = 1.3$ au, which separates near-Earth objects from the other populations.
    {Alt text: Scatter plot showing the orbital distribution of potential observable SSSBs.}
    }
    \label{fig:orbelem_cand}
\end{figure}

\begin{figure}
	\includegraphics[width=\linewidth]{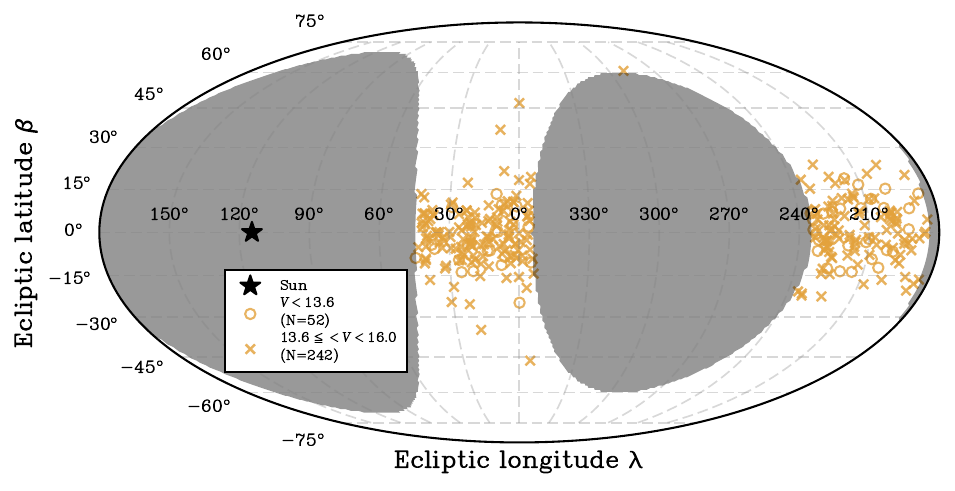}
    \caption{
    Distribution of the potential candidates on January 1, 2032, in ecliptic coordinates. 
    Potential candidates with $V < 13.6$ and $13.6 \le V < 16.0$ are represented by open circles and crosses, respectively.
    The Sun is marked with a star, and regions that are unobservable due to the solar elongation constraint are shaded.
    {Alt text: Ecliptic sky map showing potentially observable targets and the region excluded by solar elongation constraints.}
    }
    \label{fig:lambeta}
\end{figure}

A total of \Nfinalnom\ objects and \Nfinalext\ objects were extracted as nominal and extended candidates, respectively,
as shown in Fig.~\ref{fig:H_hist_1}. Their absolute magnitudes span a wide range from $H=-0.53$ for (134340)~Pluto down to $H=18.23$ for (395289)~2011~BJ$_2$ in the nominal case, and down to $H=22.14$ for (469445)~2002~LT${_{24}}$ in the extended case. 
The candidate population comprises \NMBAnom\ main-belt objects and \NNEAnom\ near-Earth objects under nominal conditions. 
Under extended conditions, this population expands to \NMBAext\ main-belt objects, \NNEAext\ near-Earth objects, and \NOtherext\ other objects, as shown in Fig.~\ref{fig:orbelem_cand}.
The other objects comprise 20 Jupiter Trojans, including (617)~Patroclus, the primary of the Patroclus-Menoetius binary system targeted for a flyby by NASA's Lucy mission \citep{Olkin2024, Levison2025}.
Figure \ref{fig:lambeta} illustrates their on-sky distribution in the ecliptic coordinate system
on January 1, 2032.
Since most of the candidates are MBAs, they are concentrated on the ecliptic plane.

\begin{figure*}
	\includegraphics[width=\linewidth]{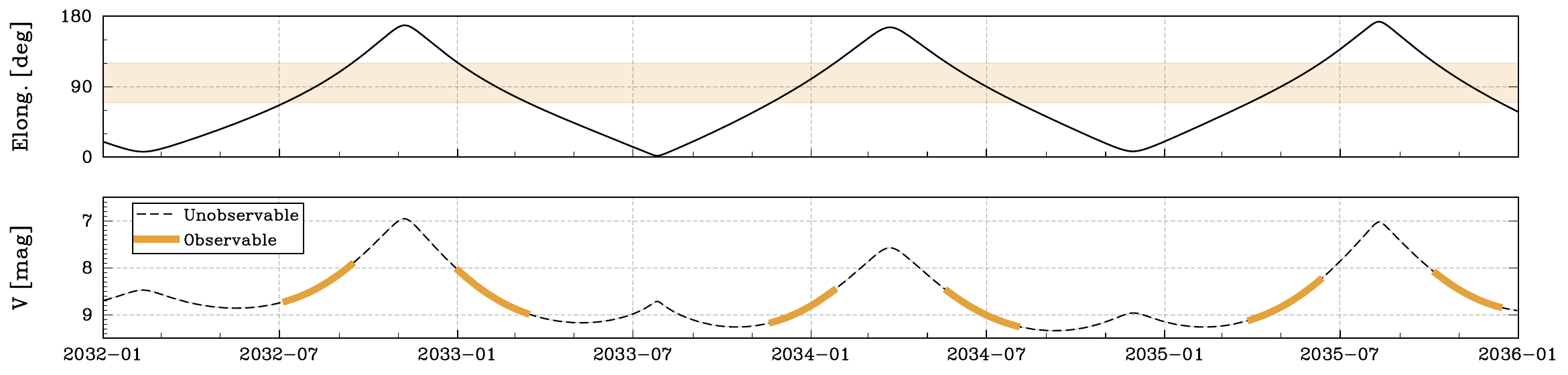}
    \caption{
    Example ephemeris of the dwarf planet in the main belt Ceres. (upper) Solar elongation. 
    The solar elongation constraints are shown as hatched.
    (lower) $V$ magnitude. 
    Periods satisfying the solar elongation constraint are shown as thick lines, while those not satisfying the constraint are shown as thin lines.
    {Alt text: Visibility plot of Ceres showing solar elongation and brightness variation over time.}
    }
    \label{fig:eph_Ceres}
\end{figure*}

\begin{figure*}
	\includegraphics[width=\linewidth]{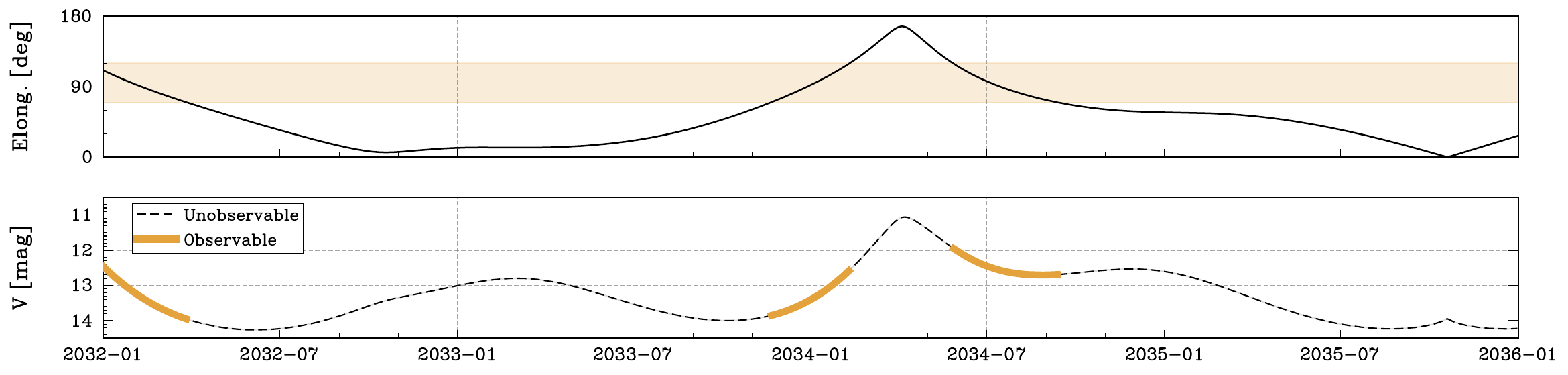}
    \caption{
    Same as Fig.~\ref{fig:eph_Ceres}, but for Eros.
    {Alt text: Visibility plot of Eros showing solar elongation and brightness variation over time.}
    }
    \label{fig:eph_Eros}
\end{figure*}

\begin{figure*}
	\includegraphics[width=\linewidth]{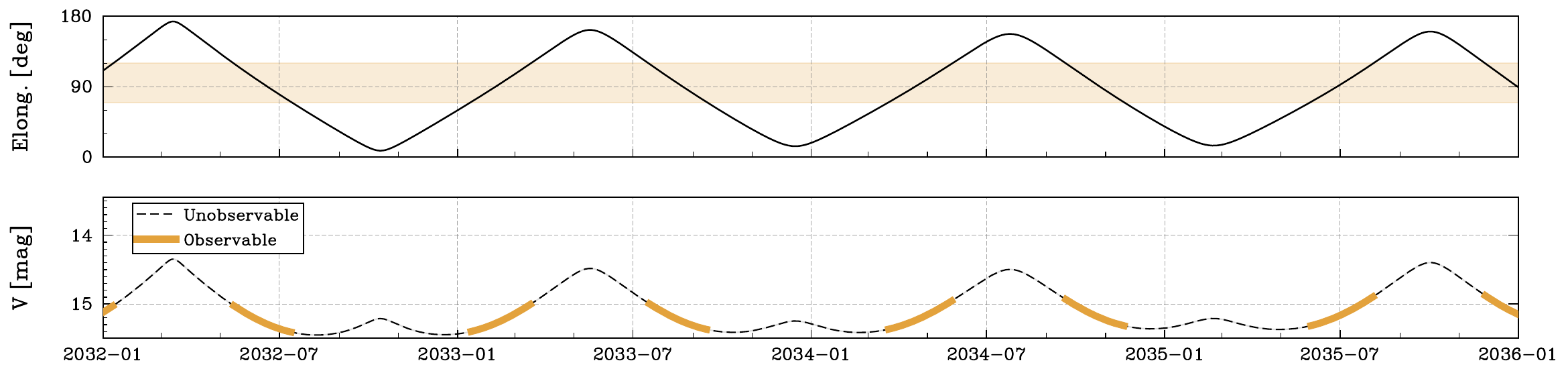}
    \caption{
    Same as Fig.~\ref{fig:eph_Ceres}, but for Hektor.
    {Alt text: Visibility plot of Hektor showing solar elongation and brightness variation over time.}
    }
    \label{fig:eph_Hektor}
\end{figure*}

Examples of ephemerides are shown in Figs. \ref{fig:eph_Ceres}--\ref{fig:eph_Hektor} for the dwarf planet in the main belt (1)~Ceres ($a=2.77$~au), the NEA Eros ($a=1.46$~au), and the Jupiter Trojan (624)~Hektor ($a=5.28$~au).
One of the brightest objects, Ceres, is always brighter than 13.6~mag and satisfies the solar elongation constraint at the six epochs.
Eros is typically brighter than 13.6 mag as well, and satisfies the solar elongation constraint at the three epochs.
Hektor is always fainter than 13.6~mag, but brighter than 16~mag, and satisfies the solar elongation constraint at eight epochs.

The number of spectral types of candidates is summarized in Table~\ref{tab:stype}.
Our candidates include all spectral types in \citet{Mahlke2022}.
Approximately 30\% of nominal targets and 20\% of extended targets are C-complex (C, Ch, and B) asteroids, which are likely to have 3-$\mu$m band absorption, although the AKARI observation showed that other spectral types sometimes exhibit absorption bands at this wavelength \citep{Usui2019}.

\begin{table}
\centering
\caption{Spectral type distribution of the Ariel candidate SSSBs$^{a}$.}
\label{tab:stype}
\begin{tabular}{crrrr}
\hline
\hline
& \multicolumn{2}{c}{Nominal} & \multicolumn{2}{c}{Extended} \\
\cline{2-3} \cline{4-5}
Type & $N$ & \% & $N$ & \% \\
\hline
S & 141 & 35.07 & 799 & 37.27 \\
C & 62 & 15.42 & 247 & 11.52 \\
Ch & 84 & 20.90 & 207 & 9.65 \\
P & 35 & 8.71 & 172 & 8.02 \\
X & 1 & 0.25 & 166 & 7.74 \\
M & 41 & 10.20 & 124 & 5.78 \\
L & 8 & 1.99 & 58 & 2.71 \\
D & 3 & 0.75 & 47 & 2.19 \\
B & 4 & 1.00 & 39 & 1.82 \\
K & 6 & 1.49 & 39 & 1.82 \\
A & 5 & 1.24 & 34 & 1.59 \\
V & 1 & 0.25 & 24 & 1.12 \\
Z & 3 & 0.75 & 18 & 0.84 \\
E & 5 & 1.24 & 17 & 0.79 \\
Q & 1 & 0.25 & 14 & 0.65 \\
T & 0 & 0.00 & 10 & 0.47 \\
R & 1 & 0.25 & 2 & 0.09 \\
O & 0 & 0.00 & 1 & 0.05 \\
Unclassified & 1 & 0.25 & 126 & 5.88 \\
\hline
\end{tabular}
 \vspace{1mm}
\parbox{\linewidth}{\footnotesize
Notes. $^{a}$ The representative taxonomy complexes were obtained using the \texttt{rocks} Python package by accessing the \texttt{rock.taxonomy.complex.value} property, which resolves and aggregates various classification schemes from the literature \citep{Berthier2023}.}
\end{table}

\begin{figure*}
	\includegraphics[width=\linewidth]{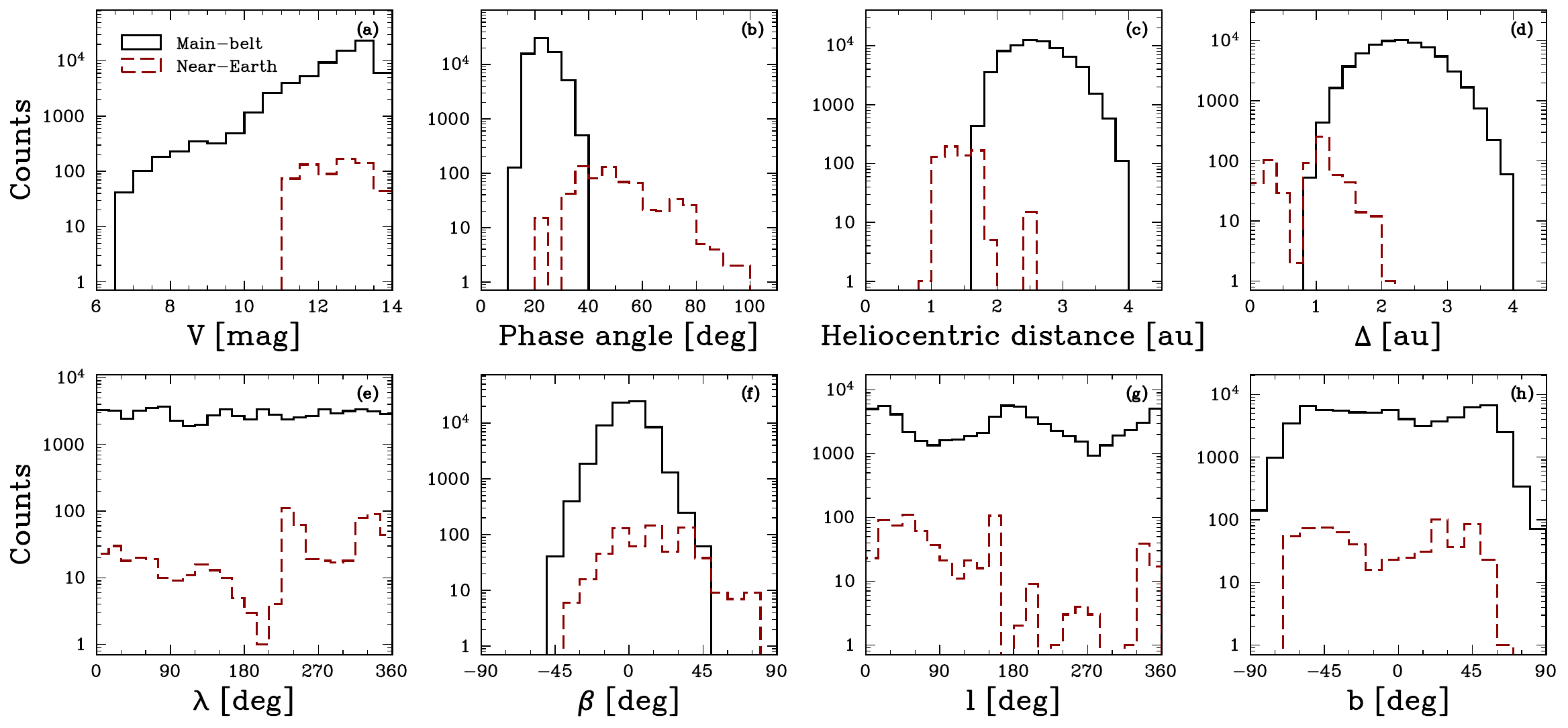}
    \caption{
    Histograms of the nominal candidates at each epoch: 
    (a) V magnitude; 
    (b) solar phase angle, $\alpha$; 
    (c) heliocentric distance, $r_h$,
    (d) observer-centric distance, $\Delta$; 
    (e) ecliptic longitude, $\lambda$; 
    (f) ecliptic latitude, $\beta$; 
    (g) galactic longitude, $l$;
    and
    (h) galactic latitude, $b$; 
    Main-belt objects and near-Earth objects
    are shown with solid and dashed 
    lines, respectively.
    {Alt text: Histograms showing the distributions of observable candidate properties, including brightness, distances, and sky positions.}
    }
    \label{fig:info_cand}
\end{figure*}

The $V$,
phase angle, 
heliocentric distances, 
and observer-centric distances,
ecliptic coordinates $(\lambda, \beta)$,
and 
galactic coordinates $(l, b)$
of nominal candidates are shown in Fig. \ref{fig:info_cand}.
The histograms include all observable epochs, and therefore contain more entries than the number of unique objects.
The candidate sample is dominated by faint objects due to the higher abundance of smaller objects.
The solar phase angle ranges are different for each dynamical class, indicating a variety in observational geometry: 
Main-belt objects typically span 10~deg to 40~deg, 
whereas near-Earth objects can reach up to 100~deg.
As also seen in Fig~\ref{fig:lambeta}, 
most of the candidates are MBAs.
Therefore, the distribution is concentrated near the ecliptic plane where these objects are most abundant.
In galactic coordinates, this concentration translates to a wide distribution where most of the objects are located within galactic latitudes of approximately $\pm 60^\circ$.

\begin{figure}
	\includegraphics[width=\linewidth]{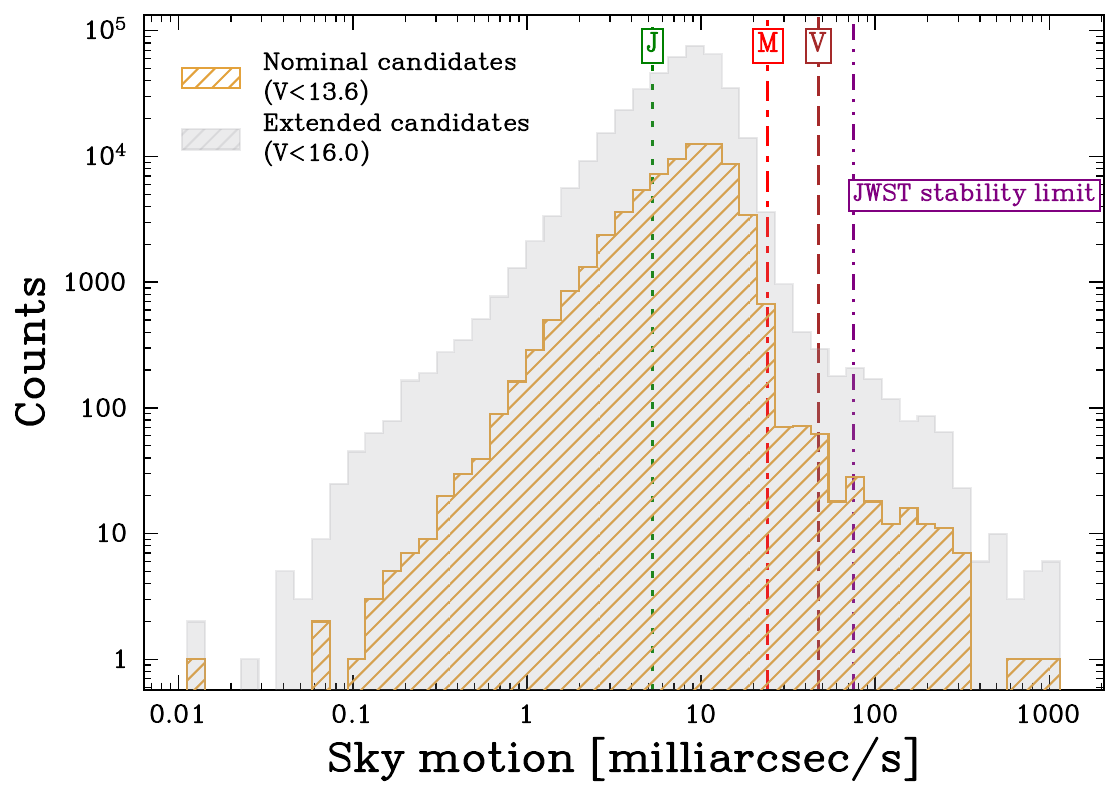}
    \caption{
    Apparent sky motion rates of Ariel potential candidates.
    The nominal candidates selected after a detailed ephemeris check with $V < 13.6$ are shown in a striped pattern, while the extended candidates with $V < 16.0$ are shown in a hatched region.
    The vertical lines indicate the typical sky motions of Jupiter (J), Mars (M), and Venus (V), 
    as well as the JWST stability limit of 75 milliarcseconds per second.
    {Alt text: Distribution of apparent sky motion rates for potential Ariel targets.}
    }
    \label{fig:skymotion}
\end{figure}

The sky motion rates of the candidates are shown in Fig.~\ref{fig:skymotion}. 
Their apparent motions depend on several factors, including $\Delta$ and the orbital elements.
Most of the candidates are MBAs, which typically have low eccentricities and are located around 1--2.5 au from Ariel. Consequently, their apparent motions are as slow as $\sim10$~milliarcsec~s$^{-1}$, significantly slower than those of the inner planets. Most targets also have apparent motions below the JWST tracking stability limit of 75~milliarcsec~s$^{-1}$
\footnote{\url{https://jwst-docs.stsci.edu/methods-and-roadmaps/jwst-moving-target-observations}}.

\begin{figure*}
	\includegraphics[width=\linewidth]{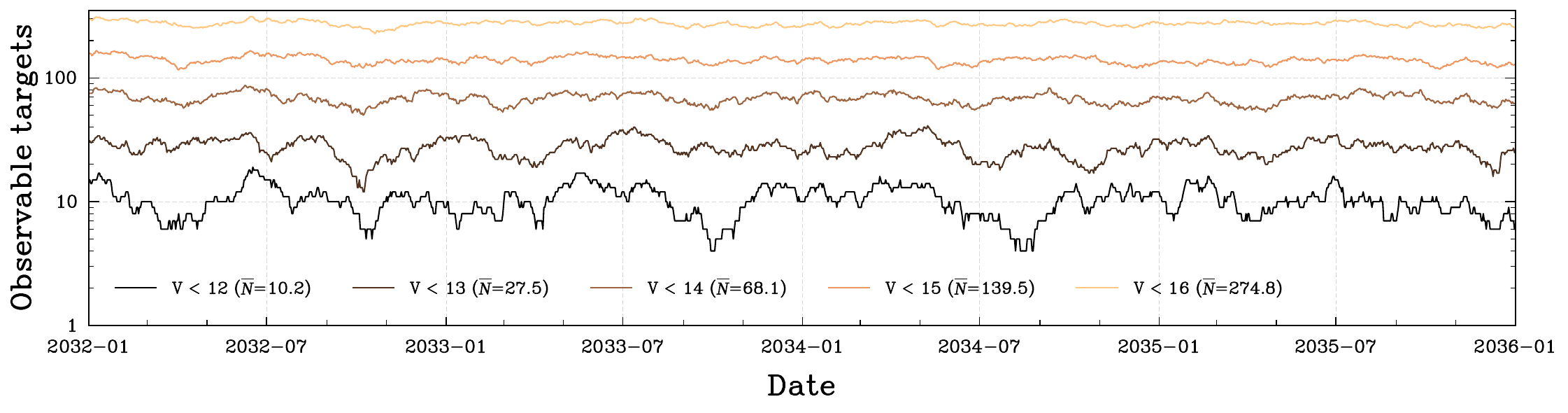}
    \caption{
    Potential candidates for various brightness levels between January 2032 and December 2035.
    The solid lines indicate the original counts sampled at 1-day intervals.
    {Alt text: Time evolution of the number of potential candidates at different brightness levels.}
    }
    \label{fig:timeseriesN}
\end{figure*}

The number of observable targets 
over time is shown in Fig.~\ref{fig:timeseriesN}.
At any time, Ariel can observe 
$\sim10$ objects with $V <12$,
$\sim30$ objects with $V <13$,
$\sim70$ objects with $V <14$,
$\sim140$ objects with $V <15$,
and
$\sim270$ objects with $V <16$.

\subsection{SEDs}
\begin{figure*}
	\includegraphics[width=\linewidth]{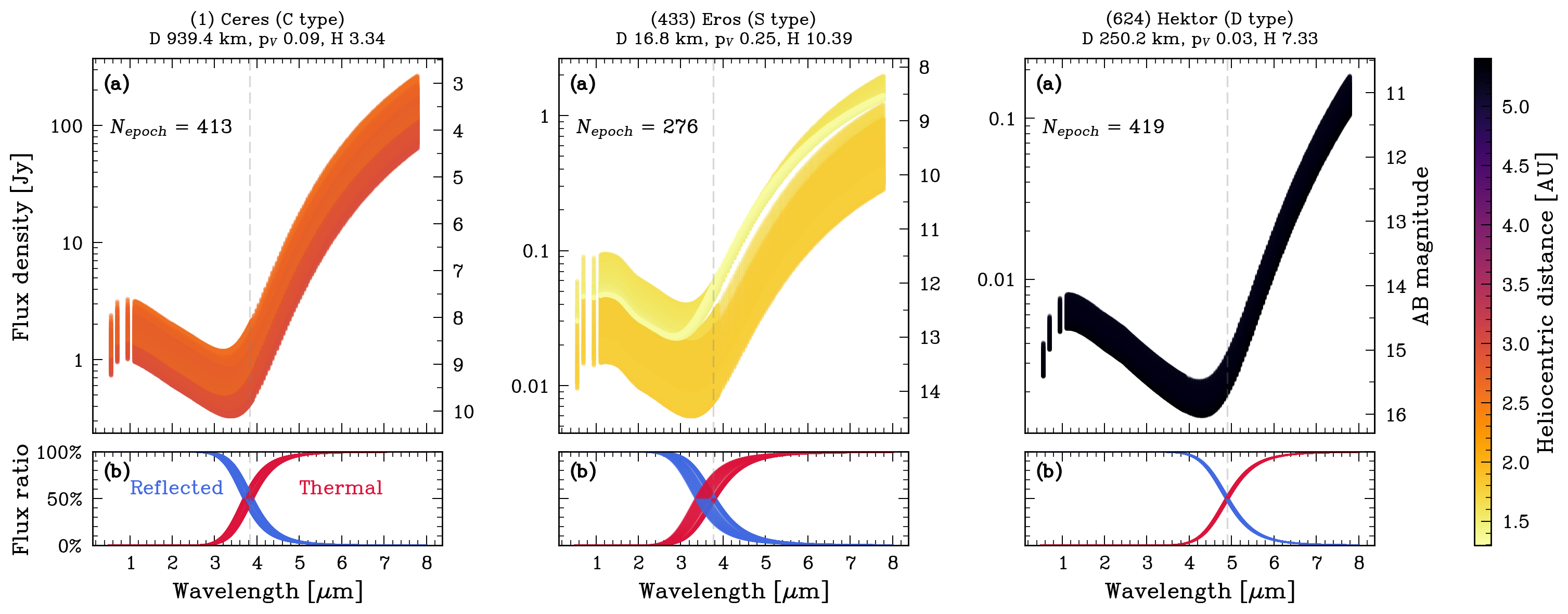}
    \caption{
        Model SED calculations for Ceres (left), Eros (middle), and Hektor (right). 
        For each target, the top panel (a) shows the predicted flux density at each epoch, while the bottom panel (b) shows the corresponding flux ratio between reflected and thermal radiation. 
        The color coding represents heliocentric distance.
        The vertical dashed lines indicate the wavelength where the reflected and thermal flux contributions are equal.
        {Alt text: Modeled spectral energy distributions of Ceres, Eros, and Hektor showing reflected and thermal emission contributions.}
        }
    \label{fig:SED}
\end{figure*}

Representative SED examples are shown in Fig.~\ref{fig:SED}.
For bright objects, we could observe them in a wide range of heliocentric distances.
For example, Ceres can be observed at heliocentric distances ranging from 2.6 to 3.0 au as shown in the left panel of Fig.~\ref{fig:SED}.
For bright NEAs, we could observe them over a wider range of heliocentric distances, and their fluxes change dramatically, 
as shown in the middle panel of Fig.~\ref{fig:SED}.
For distant objects such as Jupiter Trojans, 
the observable period is limited, and we could observe them at a limited observing geometry, as shown in the right panel of Fig.~\ref{fig:SED}.

\begin{figure*}
	\includegraphics[width=\linewidth]{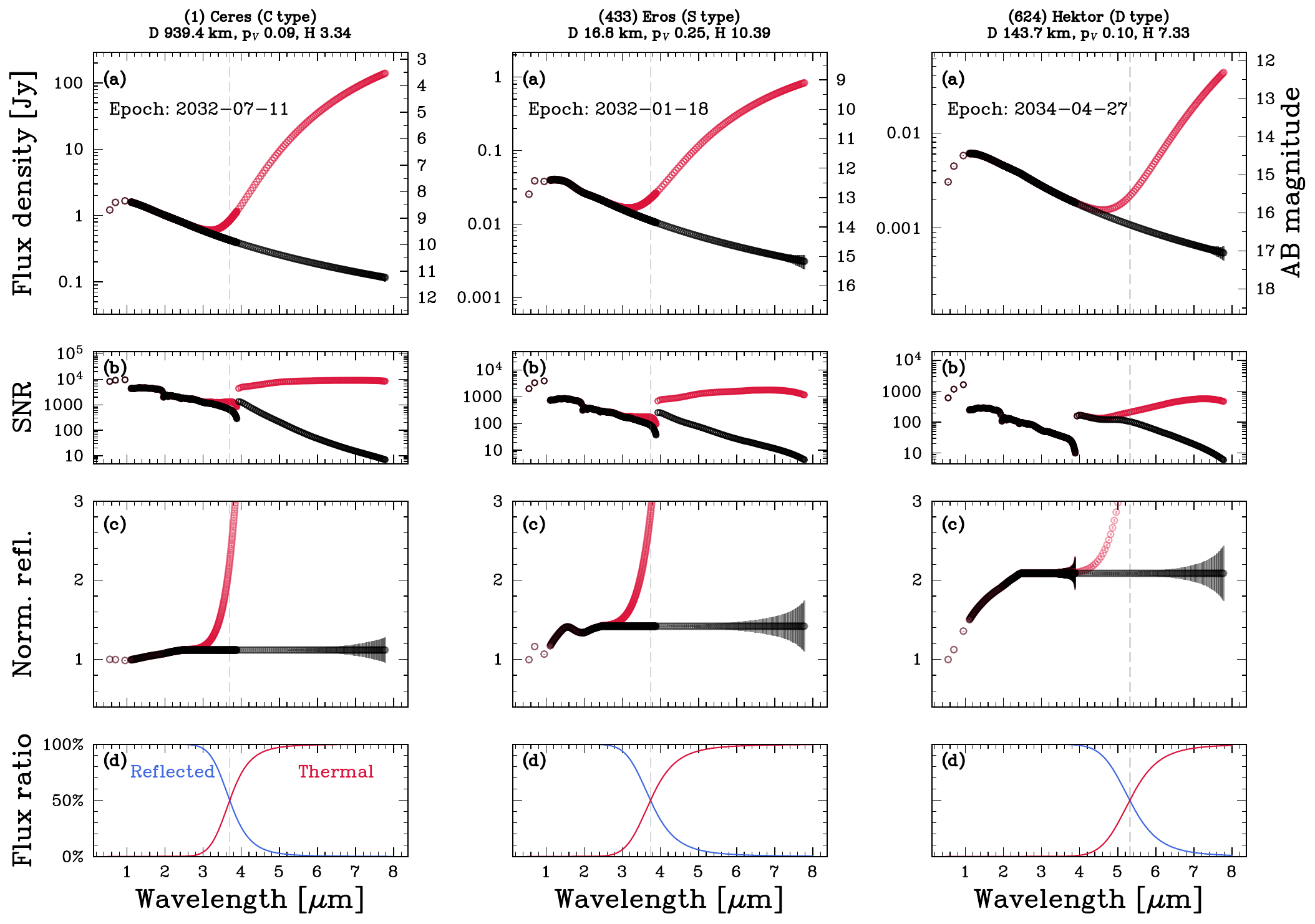}
    \caption{
    Model SED calculations for Ceres at 2032-07-11 (left), Eros at 2032-01-18 (middle), and Hektor at 2034-04-27 (right). 
        For each target, panel (a) shows the predicted flux density before and after thermal tail removal; panel (b) shows the signal-to-noise ratio (SNR) before and after thermal tail removal; panel (c) shows the normalized reflectance before and after thermal tail removal; and panel (d) shows the corresponding flux ratio between reflected and thermal radiation.
        The vertical dashed lines indicate the wavelength where the reflected and thermal flux contributions are equal.
    {Alt text: Modeled spectra of Ceres, Eros, and Hektor showing the effects of thermal tail removal on reflectance and signal-to-noise.}
        }
    \label{fig:SED_single}
\end{figure*}

Representative reflectance spectra for Ceres, Eros, and Hektor are shown in Fig.~\ref{fig:SED_single}. 
These profiles correspond to the first observable epoch for each target. For the purpose of this study, we assume that the thermal tail removal has been ideally performed.
It is clear that while the large dwarf planet Ceres and the relatively large NEA Eros exhibit high SNRs, the more distant and fainter asteroid Hektor shows a significantly lower SNR at wavelengths longer than 2~$\mu$m.

Typically, the measured fluxes in the thermal infrared have higher SNRs than those in the near-infrared wavelengths. 
However, after removing the thermal tail contribution, the derived reflectances generally exhibit lower SNRs because the thermal emission component, which contributes to the measured fluxes, is excluded.
Here, we define the transition wavelength, $\lambda_{\mathrm{trans}}$, as the wavelength at which the thermal emission equals the reflected solar flux ($F_{\mathrm{ref}} = F_{\mathrm{thermal}}$). 
For Ceres, Eros, and Hektor, the observed flux becomes dominated by thermal emission at wavelengths longer than $\lambda_{\mathrm{trans}} \approx 3.8$, $3.5$, and $5.2~\mu\mathrm{m}$, respectively.
We note that the tentative saturation limit at 7.8 $\mu$m is estimated to be approximately 70~Jy.  
An object brighter than the saturation limit will be deemed as not observable due to the risk of detector saturation and potential associated persistence effects that could compromise subsequent observations. 
Only a few bright asteroids, Ceres, 
Pallas, (4)~Vesta, and (7)~Iris are expected to exceed this limit at some times, and all of them will be observable at other times when they are less bright.

\section{Discussion} \label{sec:disc}

\subsection{Observing strategy} \label{subsec:strategy}
We estimated the number of objects observable with Ariel from January 1, 2032, to December 31, 2035, assuming that one asteroid could be targeted per complementary science window.
To perform the mock observations, we employed a selection algorithm that prioritizes the brightest available target at each epoch from the pool of currently unobserved objects.
We evaluated the total number of observable asteroids considering observation cadences of 3, 7, and 30 days.

\begin{figure*}
	\includegraphics[width=\linewidth]{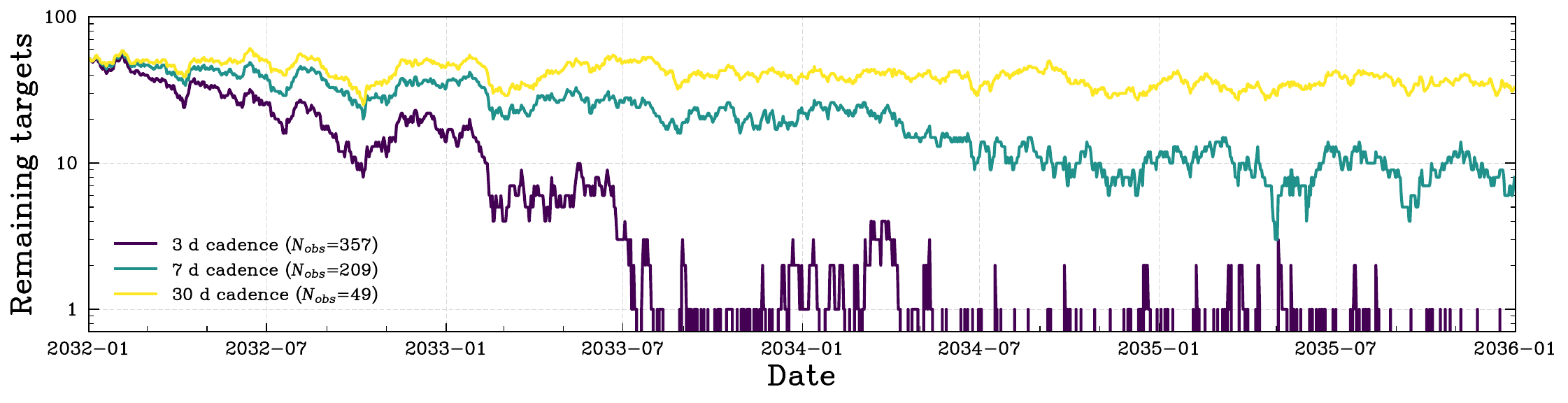}
    \caption{
    Number of remaining observable targets as a function of time for cadences of 3, 7, and 30 days.
    {Alt text: Time evolution of the number of observable targets for different observation cadences.}
    }
    \label{fig:mock}
\end{figure*}

The results of mock observations are presented in Fig. \ref{fig:mock}.
For the nominal candidates, Ariel can observe \Nthree, \Nseven, and \Nthirty~unique objects for the 3-, 7-, and 30-day cadences, respectively. 
Although the available target pool for the 3-day cadence is fully observed within 1.5 years, 
in total \Nthree~unique targets can be observed over the nominal 4-year mission.
In contrast, for the extended candidate scenario, the pool of unique targets is never depleted throughout the 4-year period across all considered cadences.
This demonstrates Ariel's potential to acquire high-quality spectra for a population of asteroids that is comparable to, or even an order of magnitude larger than, that observed by AKARI.
Given that SSSBs are bright objects widely distributed across the sky, as shown in Fig.~\ref{fig:lambeta}, they represent an ideal target population for complementary science.

\subsection{Comparison with other space missions}
\begin{figure}
	\includegraphics[width=\linewidth]{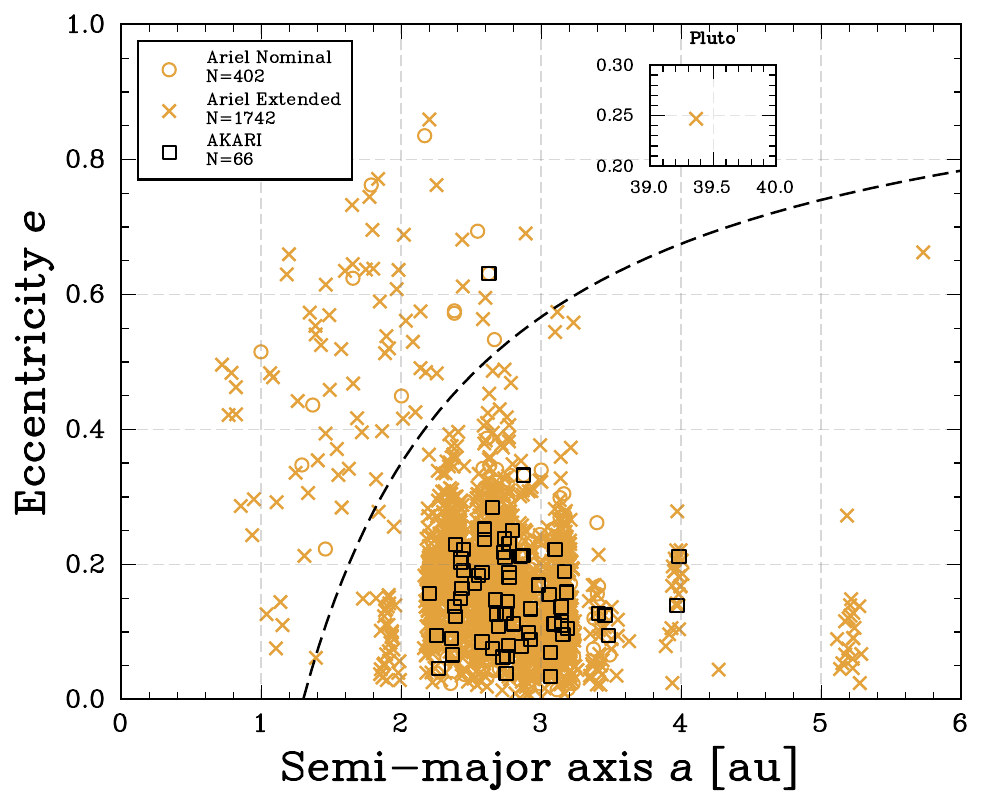}
    \caption{
    Distributions of the Ariel nominal candidates (circles), Ariel extended additions (crosses), and AKARI-observed SSSBs (squares)
    in the semi-major axis–eccentricity plane.
    The dashed line indicates the boundary corresponding to a perihelion distance of $q = 1.3$ au, which separates NEAs from the other populations.
    {Alt text: Semi-major axis–eccentricity distribution of Ariel candidates and AKARI-observed SSSBs.}
    }
    \label{fig:akarisample}
\end{figure}

The comparison with other space telescopes is summarized in Table \ref{tab:comp}.
The direction of AKARI observations was restricted to solar elongation angles within $90^\circ \pm 0.7^\circ$ due to the constraints of its orbit and attitude control system. 
Thus, the number of candidates is severely limited. 
In contrast, Ariel benefits from a broader solar elongation constraint between 70~deg and 120~deg, providing more than 400 nominal candidates and over 2,000 extended candidates, as shown in Fig. \ref{fig:H_hist_1}.
Notably, among the 66 targets observed by AKARI, 63 are included in Ariel's nominal candidate list, and all 66 targets fall within the extended candidate list, as shown in Fig.~\ref{fig:akarisample}.
Although Ariel will have a lower spectral resolution than JWST, it will enable observations of a significantly larger sample.

\begin{figure*}
    \includegraphics[width=\linewidth]{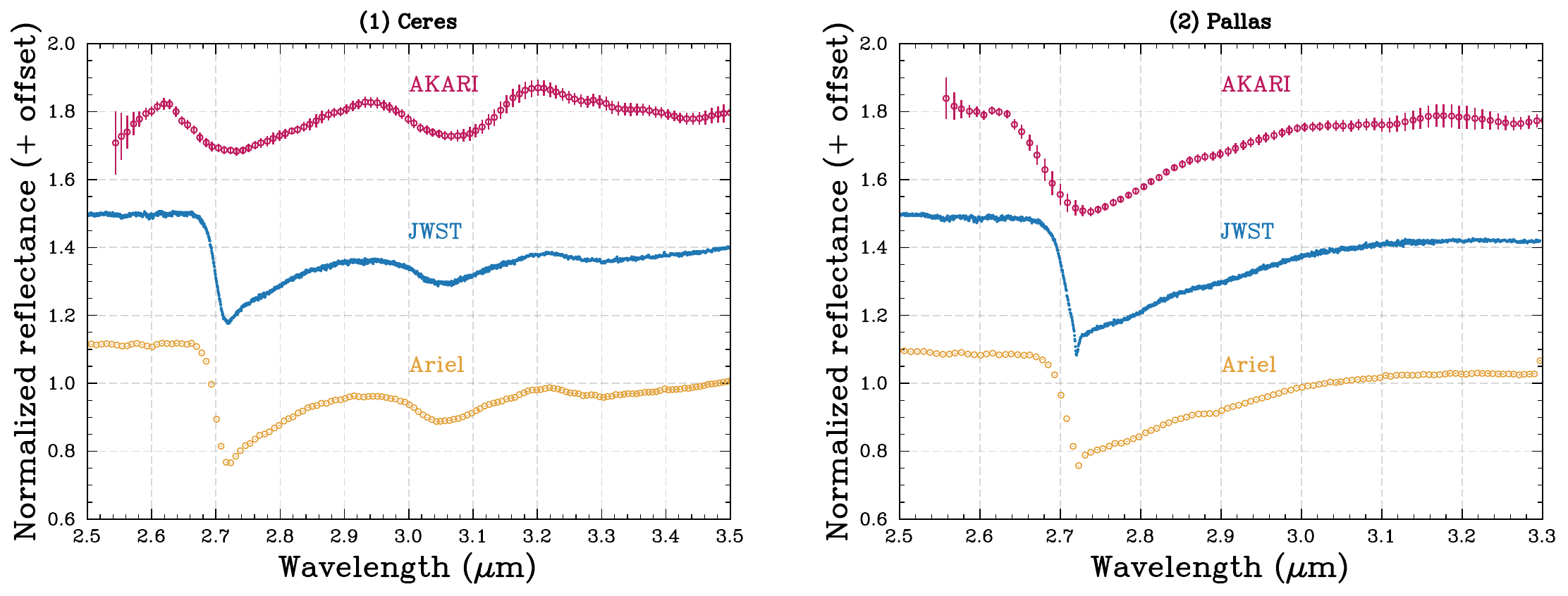}
    \caption{
    Reflectance spectra of Ceres on July 11, 2032 (left) and Pallas on November 26, 2033 (right) obtained with Ariel, compared with those from AKARI/IRC \citep{Usui2019} and JWST/NIRSpec \citep{Rivkin2025}.
    AKARI/IRC spectra are blurred due to the lack of non-sidereal tracking during the exposure.
    {Alt text: Reflectance spectra of Ceres and Pallas comparing Ariel observations with previous infrared measurements.}
    }
    \label{fig:Ceres_Pallas_3micron}
\end{figure*}

Fig.~\ref{fig:Ceres_Pallas_3micron} shows the results of simulations with spectra of Ceres and Pallas from JWST/NIRSpec \citep{Rivkin2025} and AKARI/IRC \citep{Usui2019}. 
Ariel achieves a higher spectral resolution than AKARI. 
It yields 143 data points within the 2.5–3.5 $\mu$m range for Ceres (compared to AKARI's 101 points), and 111 points within the 2.5–3.3 $\mu$m range for Pallas (compared to AKARI's 78 points).
The 2.7~$\mu$m absorption feature attributed to OH-stretch in hydrated minerals can be identified in both Ceres and Pallas spectra (see Section~\ref{subsec:outcome}). 

SPHEREx is expected to obtain spectra of more than $O(10^5)$ SSSBs for two years of survey \citep{Crill2020}.
However, only a smaller fraction of these objects, about $O(10^4)$, are expected to have high-quality data, defined here as a single-observation signal-to-noise ratio of at least 5 at 2.4 $\mu$m \citep{Ivezic2022}.
Moreover, the calibration of SPHEREx spectra for SSSBs is challenging, as the observations are not simultaneous for all wavelengths.
Obtaining science-ready reflectance spectra requires additional correction to account for changes in the target-SPHEREx geometry and the SSSB's rotation over the multi-day observation period \citep{Ivezic2022, Lim2026}.
The SOLO has been obtaining high-cadence lightcurves of SSSBs in support of the SPHEREx Solar System Object Catalog, enabling rotational variability corrections for approximately $O(10^3)$ asteroids in SPHEREx observations \citep{Lim2026}.

 
\begin{table*}
	\centering
	\caption{Comparison of NIR spectroscopy including 3 $\mu$m from space telescopes.}
	\label{tab:comp}
	\begin{tabular}{lcccccc} 
		\hline
		Instrument & Wavelength & Resolving power at 3\,$\mu$m & $N_{\rm sssb}$  & Note & Reference\\
		        & [$\mu$m]   &                         &                 &      &           \\
		\hline
		ISO/SWS   & 2.5--3.5 & $\sim100$? $^{a}$ & 2     & (1) Ceres and (2) Pallas & \citet{Rivkin1997} \\
		AKARI/IRC   & 2.5--5.0 & $\lesssim 100$ ~$^{b}$ & 66     & MBAs, Cybeles, Hildas, and one NEA                     & \citet{Usui2019} \\
        JWST/NIRSpec    & 0.6--5.3 & $\sim2700$ & $O(10)$     & As of 2026, High-resolution gratings                   & e.g., \citet{Rivkin2025, Sharkey2025} \\
		SPHEREx & 0.75--5.0 & $\sim50$ & $O(10^{3})$ & Expected for two years             & \citet{Ivezic2022, Lim2026} \\
		Twinkle & 0.40--4.5 & $\sim60$ & Not Available & Design life of seven years             & \citet{Edwards2019b} \\
		Ariel   & 0.50--7.8 & $\sim140$ & 357    & One object per day   & This study\\
        		&   &   & 209    & One object per week  & This study\\
                &   &   & 49     & One object per month & This study\\
		\hline
	\end{tabular}
 \vspace{1mm}
\parbox{\textwidth}{\footnotesize
Notes. 
$^{a}$ The resolutions are approximate values estimated from the published reflectance spectra.
$^{b}$ The effective resolving power depends on the target asteroid because no non-sidereal tracking was performed during the effective exposure time of 44.41 s per frame, resulting in image blur due to the asteroid's apparent motion.
}
\end{table*}

\subsection{Expected scientific outcomes} \label{subsec:outcome}

\subsubsection{OH feature at 2.7-$\mu$m band} \label{subsubsec:OH}
As an example, we focus on the 2.7-$\mu$m band 
among several diagnostic spectral features within Ariel’s wavelength coverage.
The 2.7-$\mu$m absorption feature is attributed to 
OH-stretch in hydrated minerals such as phyllosilicates \citep{Rivkin2015}.
We calculate the mean SNR over the 2.6--2.8~$\mu$m range for all candidate asteroids at the epoch when their $V$-band magnitude is at its median value.

The left panel of Fig~\ref{fig:snr_3micron} shows the histogram of the average SNR around 2.7 $\mu$m.
Among the nominal candidates, 392 asteroids exhibit a mean SNR of 50 or higher in the 2.7-$\mu\mathrm{m}$ region, and 307 of them reach an SNR of 100 or higher.
While prominent absorption features, such as those observed on Ceres and Pallas, can be identified even at lower SNR levels, achieving these high SNRs enables the robust evaluation of exceptionally weak features.

As shown in Fig.~\ref{fig:Ceres_Pallas_3micron}, the spectral resolution of Ariel is sufficient to resolve the structural details of the absorption features observed in Ceres and Pallas when the targets are tracked.
It is known that the OH-band minima are sensitive to the Fe/Mg inside phyllosilicates. 
And this variation is a key to constrain the aqueous alteration process among the asteroids \citep{Usui2019}.
Even for the sharper and deeper feature around 2.72~$\mu$m observed on Pallas, Ariel has the potential to capture it, although it may be resolved by only a single data point.
Moreover, there is a small absorption at 3.05~$\mu$m for Ceres, possibly ammonium-bearing phyllosilicates \citep{DeSanctis2015}. 
We can easily resolve this feature of $\sim$5\% absorption with the high SNR of Ariel.

Two NEAs out of nominal \NNEAnom~NEA samples are known to be C-complex:
Wilson-Harrington (B-type) and (7092)~Cadmus (C-type).
The 2.7-$\mu$m band of C-complex NEAs has been reported for Ryugu and Bennu by the Hayabusa2 and OSIRIS-REx missions, respectively \citep{Kitazato2019, Hamilton2019}, 
whereas no such feature was detected by JWST for (3200)~Phaethon \citep{Thomas2025}.
We could investigate the 3-$\mu$m feature of C-complex NEAs with Ariel.
Moreover, a possible 3-$\mu$m feature is reported for some S-complex NEAs in our samples such as 
Eros and (1036)~Ganymed with the NASA/IRTF \citep{Rivkin2018, McGraw2022, McGraw2024a, McGraw2026}. It is possibly because of the OH or H$_2$O production by ion irradiation \citep{Ichimura2012, Nakauchi2021}.
The expected SNRs at 2.7~$\mu$m for Eros and Ganymed are 
$\sim100$ and $\sim250$, respectively, even with the high thermal irradiation contamination, which is enough sensitivity to confirm the absorption feature of Eros reported in previous studies  \citep{Rivkin2018}.
However, the spectra of NEAs may suffer from significant
smearing, or they may be deemed unobservable,
if their sky motions exceed the currently uncertain stability limit.
While observing these targets near their stationary points provides a practical way to minimize their apparent motion \citep[e.g.,][]{Nakamura1998}, such opportunities are naturally limited.
Therefore, implementing a tracking mode capable of handling speeds up to $\sim100$~milliarcsec~s$^{-1}$ is essential to flexibly acquire reliable and scientifically meaningful spectra of NEAs.

\subsubsection{Organic feature at 3.4-$\mu$m band}
As another example, we focus on the 3.4-$\mu$m absorption band, which is associated with the presence of aliphatic hydrocarbons and/or carbonates. 
While ground-based observations are challenging, \citet{Rivkin2010} and \citet{Campins2010} reported the first detection of this feature on the MBA (24) Themis using the NASA/IRTF, followed by a similar detection on (65) Cybele \citep{Licandro2011}. 
In recent years, space-based and sample-return missions have significantly expanded our understanding; the 3.4-$\mu$m feature, resolving different types of carbonates, was detected on the NEA Bennu by OSIRIS-REx \citep{Kaplain2020}, and carbonates have been explicitly identified in returned samples from both Ryugu and Bennu \citep{Loizeau2023, Mahlke2026}. 
Furthermore, recent JWST/NIRSpec observations have revealed the presence of 3.4-$\mu\mathrm{m}$ features among Jupiter Trojans \citep{Wong2024}. 
On the other hand, \citet{LePivert-Jolivet2025} recently reported no clear detection of organic carbons for the C-complex MBA (84) Klio using JWST/NIRSpec, highlighting the potential compositional diversity among these objects. 

Beyond confirming individual detectabilities, Ariel offers a unique opportunity to address fundamental planetary science questions using the 3.4-$\mu\mathrm{m}$ band. 
Because ground-based observations in this spectral region are severely hampered by strong telluric absorption, Ariel will execute a homogeneous, space-based survey across a broad target sample. 
Crucially, Ariel enables simultaneous measurement of the 2.7-$\mu\mathrm{m}$ ($\mathrm{OH}$), 3.1-$\mu\mathrm{m}$ (water ice), and 3.4-$\mu\mathrm{m}$ (organics) absorption features. 
This capability allows for systematic investigations into the correlations among surface hydration, water ice, and organic abundances across a statistically significant sample of asteroids. 
Linking these spectral signatures will provide key insights into asteroid--meteorite relationships, the spatial distribution of volatile and carbon-bearing materials in the early Solar System, and the potential role of primitive asteroids in delivering prebiotic organic compounds to the terrestrial planets.

\begin{figure*}
	\includegraphics[width=\linewidth]{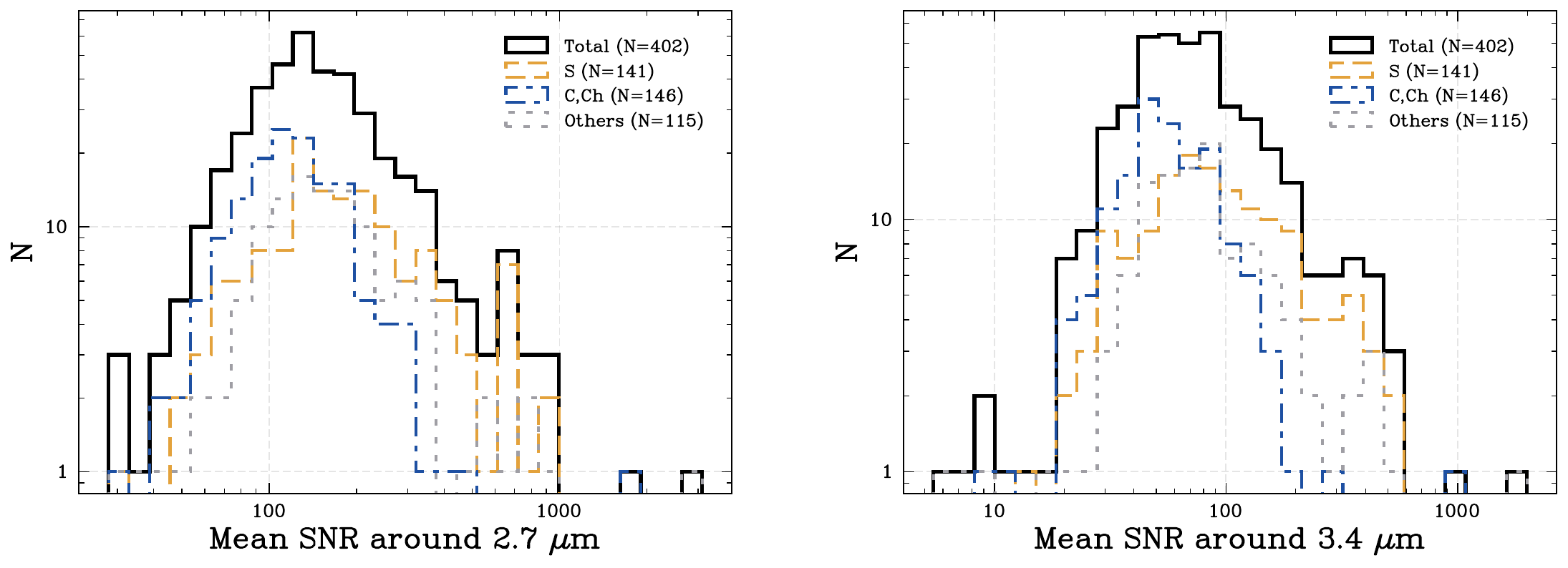}
    \caption{
     SNRs of reflectance spectra at 2.7 $\mu$m (2.6--2.8 $\mu$m) and 3.4 $\mu$m (3.3--3.5 $\mu$m).
    S, C (combined C and Ch), and others are shown as separate lines, with the total indicated by the solid line.
    {Alt text: Signal-to-noise ratios of reflectance spectra around key absorption bands for different asteroid types.}
    }
    \label{fig:snr_3micron}
\end{figure*}

To evaluate Ariel's performance for this band, we calculate the mean SNR over the 3.3--3.5~$\mu\mathrm{m}$ range using the same median V-band magnitude criterion in Section \ref{subsubsec:OH}.
The right panel of Fig.~\ref{fig:snr_3micron} shows the histogram of the average SNR around 3.4~$\mu$m. 
Among the nominal candidates, 
280 asteroids exhibit a mean SNR of 50 or higher in the 3.4-$\mu$m region, 
and 111 of them reach an SNR of 100 or higher. 
While prominent absorption features can be identified even at lower SNR levels, achieving these high SNRs ensures the robust evaluation of weak features.

In \citet{Rivkin2010}, 
a residual absorption band centered near 3.4~$\mu\mathrm{m}$ with a width of approximately 0.2~$\mu\mathrm{m}$ has been detected and attributed to complex organic materials. 
As shown in Fig.~\ref{fig:Ceres_Pallas_3micron}, the spectral resolution of Ariel is more than adequate to resolve and evaluate such a 3.4-$\mu\mathrm{m}$ absorption feature. 
Although the number of confirmed observational examples remains limited, Ariel's spectral resolution is sufficient to robustly identify features with a width of 0.1~$\mu\mathrm{m}$ or greater, as well as to resolve even narrower or more subtle sub-features within the band.

\subsubsection{Thermal properties}
In parallel with the compositional characterizations in reflectance spectra, 
Ariel's wavelength coverage allows for the constraint of thermal properties. 
Specifically, by utilizing data that extend to wavelengths longer than 3~$\mu\mathrm{m}$, it can constrain the diameter, geometric albedo, and beaming parameter \citep[e.g.,][]{Rivkin2005, Sharkey2026}.
By focusing on even longer wavelengths where thermal emission becomes dominant, it may even enable the retrieval of an emissivity spectrum \citep[e.g.,][]{Humes2024b}.

In contrast to the four-band photometry from WISE/NEOWISE \citep{Wright2010}, Ariel delivers continuous spectroscopy across the reflected-to-thermal transition region ($\sim2$--$8\,\mu\mathrm{m}$). 
This capability allows for a self-consistent separation of the reflected light and thermal emission components, substantially mitigating degeneracies between diameter, geometric albedo, and the beaming parameter. 
Although Ariel's spectral coverage does not reach the thermal emission peak near $10$--$20\,\mu\mathrm{m}$, its dense sampling along the Wien tail of the thermal continuum imposes tight constraints on surface temperature and thermal beaming. 
Consequently, Ariel observations will synergistically complement existing WISE/NEOWISE radiometric data \citep[e.g.,][]{Masiero2011, Mainzer2011a} by refining asteroid physical parameters and enhancing the calibration of thermal models across diverse taxonomic classes.

\begin{figure}
	\includegraphics[width=\linewidth]{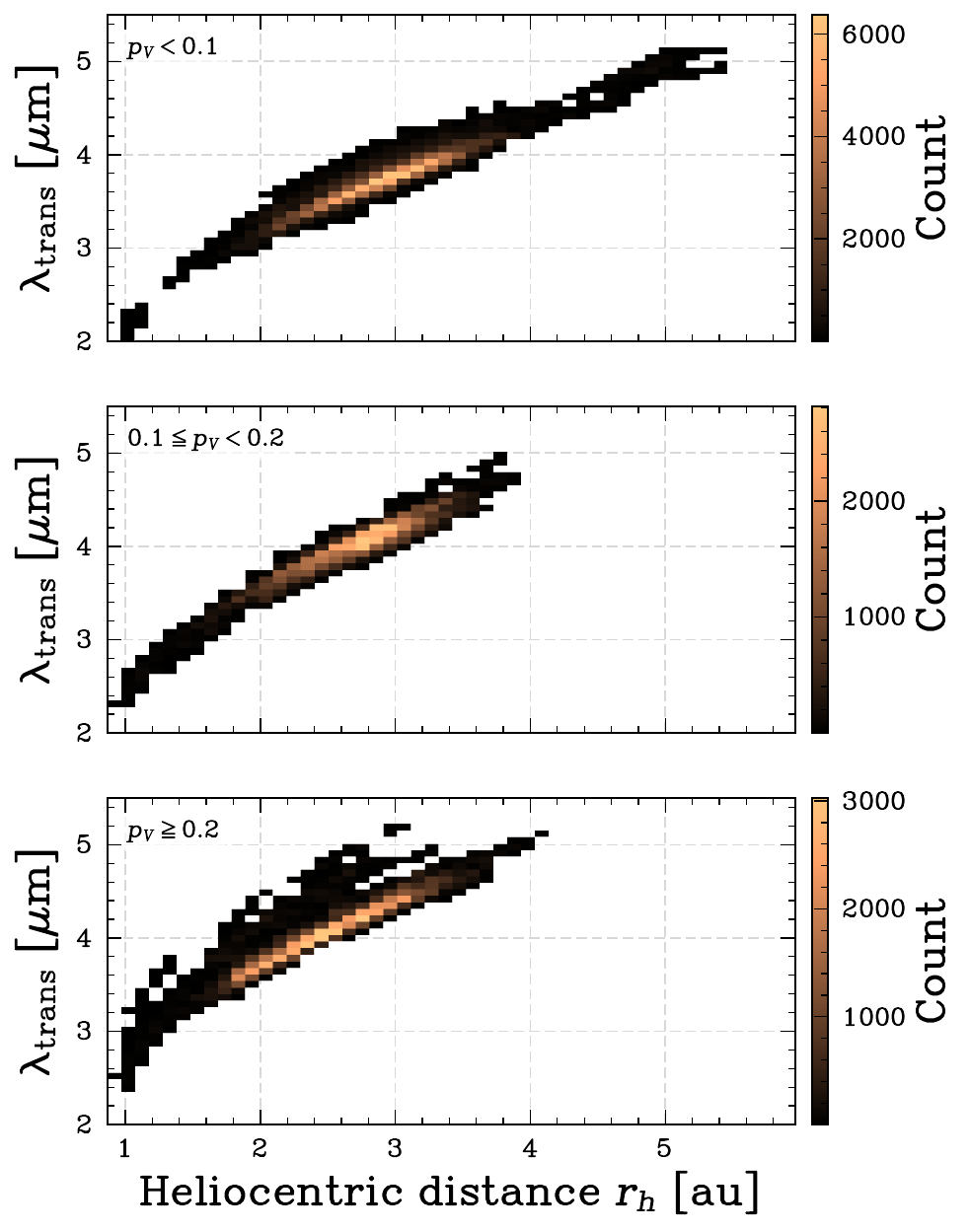}
    \caption{
    Heliocentric distance versus transition wavelength where the thermal emission equals the reflected solar flux ($F_{\mathrm{thermal}} = F_{\mathrm{ref}}$). 
    The data density for the observation epochs of candidate asteroids is presented as two-dimensional histograms across three geometric albedo ($p_V$) bins, corresponding to $p_V < 0.1$ (top), $0.1 \leq p_V < 0.2$ (middle), and $p_V \geq 0.2$ (bottom).
    The colorbar on the right of each panel indicates the number of data points (count) within each grid cell.
    {Alt text: Two-dimensional distributions of transition wavelength and heliocentric distance for candidate asteroids.}
    }
    \label{fig:FthermalFref}
\end{figure}

Figure \ref{fig:FthermalFref} illustrates the 
transition wavelength, $\lambda_\textrm{trans}$, as a function of heliocentric distance, reflecting a complex dependency on multiple factors including the albedo and spectral type.  
Specifically, this trend reflects that asteroids with larger heliocentric distances and higher albedos allow us to obtain reflectance spectra up to longer wavelengths, whereas those with smaller heliocentric distances and lower albedos enable the retrieval of emissivity spectra down to shorter wavelengths \citep[e.g.,][]{Usui2019}.

Furthermore, we assumed a fixed beaming parameter in our calculations. 
Incorporating shape models would allow for more sophisticated modeling, enabling us to employ the thermophysical model \citep[TPM, e.g.,][]{Delbo2015} to directly constrain the more physical parameter such as thermal inertia. 
However, such detailed investigations are left beyond the scope of this work.

\subsubsection{Synergy with other facilities}
Finally, organizing collaborative observations with other facilities is highly valuable, as demonstrated by several recent studies. 
For instance, \textit{SPHEREx} observations are being closely coordinated with ground-based lightcurve measurements to properly account for the brightness variations of asteroids caused by their rotation \citep{Lim2026}.

To maximize the scientific output of the observations of SSSBs with Ariel, we plan to organize collaborative observations with other assets. 
Specifically, quasi-simultaneous observations using ground-based telescopes will be conducted to obtain data in the near-ultraviolet (NUV) and 0.7~$\mu\mathrm{m}$ regions, providing critical complementary information \citep{Takir2012, Tatsumi2023, Tinaut-Ruano2024, Tatsumi2025}. 
Furthermore, polarimetric observations are crucial to constrain the surface properties of hydrated asteroids, such as porosity and particle size \citep{Geem2024}.
Through this observational strategy, we will obtain complete reflectance spectra spanning from the NUV to the NIR for several hundred asteroids, which will dramatically advance our statistical understanding of the compositional distribution of small Solar System bodies.

\section{Conclusions}
This study demonstrates that the ESA Ariel space mission holds significant potential for the spectral characterization of SSSBs within its 
0.5--7.8~$\mu$m wavelength range. 
Our visibility analysis reveals that Ariel can observe up to \Nfinalnom~unique SSSBs under the nominal visual magnitude limit ($V=13.6$), which increases to \Nfinalext~unique SSSBs under the extended visual magnitude limit ($V=16.0$).
Notably, due to their very slow sky motions ($\sim 10\ \text{milliarcseconds}\ \text{s}^{-1}$), 
these SSSBs are highly flexible and competitive targets within the complementary science program.
Spectral energy distribution modeling confirms that Ariel will successfully resolve key absorption features at 
2.7~$\mu$m and 3.4~$\mu$m with superior spectral resolution compared to past missions like AKARI. 
These observations will provide unprecedented insights into the distribution of hydrated materials and organics, securing SSSB science as a compelling component of the Ariel mission.

\section*{Acknowledgements}
We thank Andy Rivkin for providing the JWST/NIRSpec spectra of the asteroids.
We thank the reviewers for their constructive comments.
This work was supported by JSPS KAKENHI Grant Numbers JP22K21344, 25H00665, and 26K07174.
This work was supported by the French government through the France 2030
investment plan managed by the National Research Agency (ANR), as part of the
Initiative of Excellence Université Côte d’Azur under reference number ANR-15-IDEX-01.
This work was supported by the Japan Society for the Promotion of Science (JSPS) Overseas Research Fellowships.
J.B. acknowledges support from the IAC Early-Career Visitor Programme.
E.T. acknowledges support from Spanish government - Ayuda Ram\'{o}n y Cajal 2023 (RYC2023-043483-I).
SH was supported by the Hypervelocity Impact Facility (former name: The Space Plasma Laboratory), ISAS, JAXA.
M.G., A.P., and L.V.M. acknowledge grant support from the UK Space Agency.
C.K. received funding from the PRODEX programme of the Europen Space Agency under the contract PEA 4000149210.
The spectral types and albedo of asteroids were retrieved thanks to the \texttt{SsODNet} service
\citep{Berthier2023}
of the Space Service (SE-OP)
of Laboratoire Temps Espace at Paris Observatory through \texttt{rocks} Python interface\footnote{\url{https://rocks.readthedocs.io/en/latest/}}.


\section*{Conflict of Interest}
The authors declare no conflict of interest.

\section*{Data Availability}
The data underlying this article will be shared on reasonable request to the corresponding author.
\bibliographystyle{rasti}
\bibliography{paplog}

\bsp	
\label{lastpage}
\end{document}